\RequirePackage{amsmath}
\documentclass[runningheads,envcountsame]{llncs}
\usepackage{amssymb}
\usepackage{verbatim}
\usepackage{gastex}
\usepackage{algorithmicx,algorithm,algpseudocode}
\usepackage{pstricks,pstricks-add,pst-node}
\usepackage{array,graphicx,tabularx,multirow}
\usepackage{xcolor}
\usepackage[lowtilde]{url}

\spnewtheorem{conjecture}{Conjecture}{\bfseries}{\rmfamily}
\spnewtheorem{pbm}[conjecture]{Problem}{\bfseries}{\rmfamily}

\usepackage[utf8]{inputenc}
\usepackage[T1]{fontenc}
\DeclareSymbolFont{rsfscript}{OMS}{rsfs}{m}{n}
\DeclareSymbolFontAlphabet{\mathrsfs}{rsfscript}

\newcommand{\Q}{\mathbb{Q}}

\begin{document}
\title{Experiments with Synchronizing Automata}
\author{Andrzej Kisielewicz\thanks{Supported in part by the National Science Centre, Poland under project number 2012/07/B/ST1/03318.}
\and Jakub Kowalski\thanks{Supported in part by the National Science Centre, Poland under project number 2015/17/B/ST6/01893.}
\and Marek Szyku{\l}a\thanks{Supported in part by the National Science Centre, Poland under project number 2013/09/N/ST6/01194.}}
\institute{Department of Mathematics and Computer Science, University of Wroc{\l}aw
\email{andrzej.kisielewicz@math.uni.wroc.pl,\ jko@cs.uni.wroc.pl,\ msz@cs.uni.wroc.pl}}
\maketitle

\begin{abstract}
We have improved an algorithm generating synchronizing automata with a large length of the shortest reset words. This has been done by refining some known results concerning bounds on the reset length. Our improvements make possible to consider a number of conjectures and open questions concerning synchronizing automata, checking them for automata with a small number of states and discussing the results. In particular, we have verified the \v{C}ern\'y conjecture for all binary automata with at most 12 states, and all ternary automata with at most 8 states.
\end{abstract}

\section{Introduction}

A deterministic finite automaton $\mathcal{A}$ is $\langle Q,\Sigma,\delta \rangle$, where $Q$ is the set of the states, $\Sigma$ is the input alphabet, and $\delta\colon Q \times \Sigma \to Q$ is the (complete) transition function.
Throughout the paper, by $n$ we denote the number of states $|Q|$. If $|\Sigma|=k$ then $\mathcal{A}$ is called $k$-\emph{ary}.
 The transition function $\delta$ is naturally extended to a function $2^Q \times \Sigma^* \to 2^Q$.
The \emph{image} of $S \subseteq Q$ under the action of a word $w \in \Sigma^*$ is $Sw = \{\delta(q,w) \mid q \in S\}$.
The \emph{rank} of a word $w \in \Sigma^*$ is $|Qw|$, and the \emph{rank} of $\mathcal{A}$ is the minimal rank of a word over $\mathcal{A}$.
For a non-empty subset $\Sigma'\subseteq \Sigma$, we may define the automaton $\mathcal{A}' = \langle Q,\Sigma',\delta' \rangle$, where $\delta'$ is the natural restriction of $\delta$ to $\Sigma'$. In such a case $\mathcal{A}$ is called an \emph{extension} of $\mathcal{A}'$. 
The automata of rank 1 are called \emph{synchronizing}, and each word $w$ with $|Qw|=1$ is called a \emph{synchronizing} (or \emph{reset}) word for $\mathcal{A}$. An automaton is \emph{irreducibly synchronizing} if it is not an extension of a synchronizing automaton over a smaller alphabet.

We are interested in the length of a shortest reset word for $\mathcal{A}$ (there may be more than one word of the same shortest length). We call it the \emph{reset length} of $\mathcal{A}$.
The famous \v{C}ern\'y conjecture states that every synchronizing automaton $\mathcal{A}$ with $n$ states has a reset word of length $\le (n-1)^2$ \cite{Cerny1964}. This conjecture was formulated by \v{C}ern\'y in~1969 and is considered the longest-standing open problem in combinatorial theory of finite automata.
So far, the conjecture has been proved only for a few special classes of automata, and a cubic upper bound $(n^3-n)/6-1$ \cite{Pin1983OnTwoCombinatorialProblems} has been established, which was not improved for over 30 years (see \cite{KariVolkov2013Handbook,Volkov2008Survey} for excellent surveys).
The bound $(n-1)^2$ is met for every $n$ by the \v{C}ern\'y automata \cite{Cerny1964}, which is the only known infinite series of automata meeting this bound (besides that, there are 8 known particular examples with $n \le 6$ states \cite{Tr2006Trends} also meeting the bound).

There were several efforts to check computationally the conjecture for all automata with a small number of states. In particular, Ananichev, Gusev, and Volkov \cite{AGV2010,AGV2013} have checked all binary automata with at most $n=9$ states, and the checking for all automata with at most $n=10$ states was reported in \cite{Tr2006Trends}. In~\cite{KS2013GeneratingSmallAutomata}, using a dedicated algorithm, we have verified the conjecture for all binary automata with $n\le 11$ states.

In this paper, first we describe improvements to our algorithm from~\cite{KS2013GeneratingSmallAutomata}, which are aimed at making possible verifying the conjectures for larger automata. While these are results of a rather technical nature, and may be not very interesting from theoretical point of view, they make possible to restrict the computation process to much smaller class of relevant automata, and thus to consider also automata with a larger number of states.

We extend verification of the \v{C}ern\'y conjecture up to $12$ states and present an extensive experimental study on important problems and conjectures closely related to upper bounds on reset lengths.
We consider known conjectures, and restate or state new ones basing on our experiments.
Most of them imply an improvement for the general cubic bound, and hence are very hard but stand as possible ways to attack the main problem.
All of the conjectures are experimentally confirmed for automata with a small number of states and/or letters.

\section{Reset Lengths of Extensions}

In this section we describe two theoretical results we apply in the improved algorithm.
We are interested mainly in estimating the reset length of synchronizing automata that arise as extensions of non-synchronizing automata by one letter. 
In some cases, we are able to provide better upper bounds than the general bound $(n^3-n)/6-1$ \cite{Pin1983OnTwoCombinatorialProblems}.

In particular, we search for synchronizing automata with relatively large reset length. We improve the algorithm from \cite{KS2013GeneratingSmallAutomata} which takes a set of $(k-1)$-ary automata with $n$ states and generates all their nonisomorphic one-letter extensions. To perform an exhaustive search over the $k$-ary automata with $n$ states with some property, we need to progressively run the algorithm $k-1$ times starting from the complete set of non-isomorphic unary automata. However, in each run, if we know that any extension of an automaton $\mathcal{A}$ cannot have the desired property, we can safely drop $\mathcal{A}$ from further computations. Since the number of generated automata grows rapidly, suitable knowledge saves a lot of computational time and extends the class of the automata investigated.
The technical details of the algorithm and proofs can be found in \cite{KS2016GeneratingSynchronizingAutomataWithLargeResetLengths}.

A subset $M \subseteq Q$ of the states is called \emph{compressible}, if there is a word $w$ such that $|Mw| < |M|$.
Let $\mathcal{A} = \langle Q,\Sigma,\delta \rangle$ be a finite automaton. 
We say that a sequence $(M_i,x_i,y_i)$, $(1 \le i \le \ell)$ of $m$-subsets (subsets of size $m$) $M_i$ of $Q$ and pairs of states $x_i,y_i \in Q$ is an \emph{$m$-subset Frankl-Pin sequence} if the following conditions are satisfied 
\begin{enumerate}
\item $x_i,y_i \in M_i$ for $1 \le i \le \ell$;
\item either $x_i$ or $y_i$ is not in $M_j$ for all $1 \le j < i \le \ell$.
\end{enumerate}
If all the pairs $\{x_i,y_i\}$ belong to a set $P$ of pairs, we will say that this sequence is \emph{over} $P$.
Given a set $P$ of \emph{compressible} pairs, by the \emph{synchronizing height} $h(P)$ of $P$ we mean the minimal $h$ such that for each pair $\{x,y\}\in P$ there exists a word $w$ of length $h$ such that $xw=yw$. 

It is known that a shortest word compressing $M$ cannot be longer than the length of the Frankl-Pin sequence starting from $M$ \cite{Fr1982} (this, in fact, is used to obtain the bound $(n^3-n)/6$ mentioned above). Our first technical improvement is that if the synchronizing height is smaller than the maximal length of a Frankl-Pin sequence over $P$, then we have
\begin{theorem}\label{thm:pairs_bound}
Let $P$ be a set of compressible pairs in $\mathcal{A}$, $h(P)$ the synchronizing height of $P$, and $p(P)$ the maximal length of a Frankl-Pin sequence over $P$. 
Then, for every compressible $m$-subset $M$ of $Q$ ($2 \le m \le n$), there is a word compressing $M$ whose length does not exceed 
$$\binom{n-m+2}{2} - p(P) + h(P).$$
\end{theorem}

This result improves the estimation in \cite{Fr1982} by the negative summand ($p(P) - h(P)$). It is to be combined and compared with the result by J.-E. Pin \cite{Pin1972Utilisation} saying that if $w$ is a word of rank $r$ and there exists a word of rank $\le r-1$, then there is such a word of length $\le 2|w|+n-r+1$.
There are other results of this kind that can be used for providing bounds for extensions, as that in \cite{BS2015AlgebraticSynchronizationCriterion}.
Unfortunately, for small values of $n$ that are within our considerations, this does not overcome the bound from Theorem~\ref{thm:pairs_bound}.

Recall that an automaton $\mathcal{A} = \langle Q,\Sigma,\delta \rangle$ is \emph{one-cluster}, if it has a letter $a\in\Sigma$ such that for every pair $q,s \in Q$ there are $i,j \ge 1$ such that $qa^i = sa^j$. This means that the graph of the transformation induced by $a$ is connected. In particular, it has a unique cycle $C \subseteq Q$ with the property $Ca^i = C$ for every $i\ge 0$, and there is $\ell\ge 0$ such that $Qa^\ell = C$. The least such $\ell$ is called the \emph{level} of $\mathcal{A}$. 
Steinberg \cite{Steinberg2011OneClusterPrime} proved that if the length $m$ of the cycle is prime, then the one-cluster automaton $\mathcal{A}$ has a reset word of length at most
\begin{equation}\label{bounds-eq:prime}
n - m + 1 + 2\ell + (m - 2)(n + \ell).
\end{equation}
We generalize this result to arbitrary lengths and get an additional negative summand. We refine the proof of Steinberg \cite{Steinberg2011OneClusterPrime} and the summand is expressed in algebraic terms of the proof.  
Therefore, to present the result we have to recall basic notations from~\cite{Steinberg2011OneClusterPrime}. 

Given a one cluster automaton with the notation as above, we consider the matrix representation
$\pi\colon \Sigma^* \to M_n(\Q)$ defined by $\pi(w)_{q,r} = 1$ if $qw=r$, and $0$, otherwise. Given $S\subseteq Q$ we define $[S]$ to be the characteristic row vector of $S$ in $\Q^n$, $[S]^T$ its transpose, and $\gamma_S = [S]^T - (|S|/|C|)[Q]^T$.
By $w\gamma_S$ we denote the product of corresponding matrices; in particular, 
the word $w$ represents the matrix $\pi(w)$, and the product is a vector in the space $\Q^n$. We consider 
the subspace $W_S = \mbox{\rm Span}\{a^{\ell+j}\gamma_S \in \Q^n \mid 0\le j \le m-1\}$ (cf.~\cite{Steinberg2011OneClusterPrime}), and 
the \emph{cyclic period} $q_S$ of $S$, understood as the least number $q$ such that $Sa^q=S$. Now, we
define ${D}^*(m,k)$ to be the minimal value of $m-q_S+\dim W_S$ taken over all vectors $S$ with $|S|=k$. Then we prove the following:
\begin{theorem}\label{thm:one-cluster_bound}
Let $\mathcal{A} = \langle Q, \Sigma, \delta \rangle$ be a synchronizing automaton with $n$ states, such that there exists a word $w$ of length $s$ inducing a one-cluster transformation with level $\ell$ and cycle $C$ of length $m>1$.
Then $\mathcal{A}$ has a reset word of length at most
$$s(\ell +m-2)(m-1) + (n+1)(m-1)+s\ell - \sum_{k=1}^{m-1} {D}^*(m,k).$$
\end{theorem}

One can demonstrate that this results generalizes and improves earlier bounds in~\cite{CarpiDAlessandro2013IndependendSetsOfWords,Steinberg2011OneClusterPrime}, and a careful estimation of the summand $D^*(m,k)$ yields the currently best general bound for reset lengths of one-cluster automata:
\begin{corollary}\label{cor:one-cluster_estimation}
A synchronizing one-cluster automaton $\mathcal{A}$ with $n$ states and the cycle of length $m$ has a reset word of length at most
\begin{equation}
2nm-4m\ln\frac{m+3}{2}+2m-n+1
\end{equation}
\end{corollary}
Nevertheless, for small values of $m$ we can compute the exact values of $D^*(m,k)$, and this yields considerably better bounds than the general estimation above.

\section{Experiments and Conjectures}

In this section we discuss the results of our experiments with the improved algorithm concerning various conjectures and open problems in the area.

\subsubsection{The \v{C}ern\'y conjecture.}

We have verified the \v{C}ern\'y conjecture for several cases.
In particular, we confirmed it for all binary automata with $n \le 12$ states, and for all ternary automata with $n \le 8$.

Verifying the \v{C}ern\'y conjecture for binary automata with $n=12$ states was the most difficult computation that we have performed here.
The total time of a single processor core spent for this computation was about 100 years.
We performed this on a grid in parallel using mostly about 200 cores of Quad-Core AMD Opteron(tm) Processor 8350, 2.0 GHz.
The total number of automata generated by our algorithm in this case was about $10^{15}$.

For ternary automata with $n=8$ states the computation took $1.25$ years of a single processor core, and we had to generate and check about $2.1 \times 10^{10}$ automata.
One may compare these numbers with the numbers of non-isomorphic initially connected automata that one would need to generate applying the technique described in~\cite{AGV2010}. The corresponding numbers are: about $2.2 \times 10^{17}$ for binary automata with $n=12$ states, and $5.7 \times 10^{17}$ for ternary automata with $n=8$ states.

Within the range we have considered, the only automata meeting the bound $(n-1)^2$ other than the \v{C}ern\'y series are known examples with $n \le 6$ states that were presented in \cite{Tr2006Trends}.

\subsubsection{Slowly synchronizing automata.}

For the case of binary automata $n = 12$ states, we have obtained also the complete list of strongly connected synchronizing automata with reset length $\ge 94$.

\renewcommand{\arraystretch}{1.5}
\begin{table}[ht]
\centering\scriptsize
\caption{The numbers of all non-isomorphic strongly connected synchronizing binary automata with $12$ states with reset length $\ge 94$.}\label{tab:cerny12}
\begin{tabular}{|l|c|c|c|c|c|c|c|c|c|c|c|c|c|c|c|c|c|c|c|c|c|c|c|c|c|c|c|c|}\hline
Reset length                         & 94&95--98&99&100&101&102&103--109&110&111&112&113--120&121\\ \hline
Number of automata                   & 3 & 0    & 3& 21& 9 & 2 & 0      & 2 & 1 & 1 & 0      & 1\\ \hline
Series                                  &   &      &  &   &   &$\mathrsfs{H}_n$,$\mathrsfs{\dot{H}}_n$&
&$\mathrsfs{E}_n$,$\mathrsfs{D''}_n$&$\mathrsfs{W}_n$&$\mathrsfs{D'}_n$&        & $\mathrsfs{C}_n$\\ \hline
\end{tabular}
\end{table}

Table~\ref{tab:cerny12} shows the exact numbers of automata in this range, and the corresponding series according to naming from \cite{AGV2010,AGV2013,KS2013GeneratingSmallAutomata}.
Here, all automata with reset length $\ge 99$ has a similar structure of one long cycle and a small gadget (cf.~\cite{AGV2013}), and they can be generalized to series of length $n^2-O(n)$ as well.
We confirm, for $n \le 12$, \cite[Conjecture~1]{AGV2013}, which is a generalization of the \v{C}ern\'y conjecture, describing all binary synchronizing automata with reset length $\ge n^2-4n+8$ (104 for $n=12$) and stating that up to isomorphism this list is complete.

As observed in \cite{AGV2013,KS2013GeneratingSmallAutomata,Tr2006Trends}, there are gaps in the set of possible reset lengths near the \v{C}ern\'y bound $(n-1)^2$.
We confirm for binary automata that for $n=6,7,8$ there is one gap, for $n=9,10$ there are two gaps, and for $n=11,12$ there are three gaps.

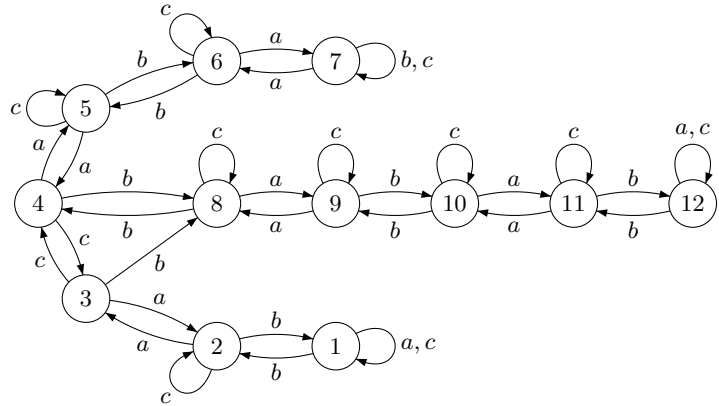
\begin{figure}[ht]
\unitlength 4.5pt
\begin{center}\begin{picture}(55,27)(0,3)
\gasset{Nh=4,Nw=4,Nmr=2,ELdist=0.5,loopdiam=3}
\node(1x)(25,4){$1$}
\node(1)(15,4){$2$}
\node(2)(4,8){$3$}
\node(3)(0,16){$4$}
\node(4)(4,24){$5$}
\node(5)(15,28){$6$}
\node(5x)(25,28){$7$}
\node(6)(15,16){$8$}
\node(7)(25,16){$9$}
\node(8)(35,16){$10$}
\node(9)(45,16){$11$}
\node(9x)(55,16){$12$}
\drawedge[curvedepth=1](1x,1){$b$}
\drawedge[curvedepth=1](1,1x){$b$}
\drawedge[curvedepth=1](1,2){$a$}
\drawedge[curvedepth=1](2,1){$a$}
\drawedge[curvedepth=1](2,3){$c$}
\drawedge[curvedepth=1](3,2){$c$}
\drawedge[curvedepth=1](3,4){$a$}
\drawedge[curvedepth=1](4,3){$a$}
\drawedge[curvedepth=1](4,5){$b$}
\drawedge[curvedepth=1](5,4){$b$}
\drawedge[curvedepth=1](5,5x){$a$}
\drawedge[curvedepth=1](5x,5){$a$}
\drawedge[curvedepth=1](3,6){$b$}
\drawedge[curvedepth=1](6,3){$b$}
\drawedge[ELdist=-2](2,6){$b$}
\drawedge[curvedepth=1](6,7){$a$}
\drawedge[curvedepth=1](7,6){$a$}
\drawedge[curvedepth=1](7,8){$b$}
\drawedge[curvedepth=1](8,7){$b$}
\drawedge[curvedepth=1](8,9){$a$}
\drawedge[curvedepth=1](9,8){$a$}
\drawedge[curvedepth=1](9,9x){$b$}
\drawedge[curvedepth=1](9x,9){$b$}
\drawloop[loopangle=0](5x){$b,c$}
\drawloop[loopangle=225](1){$c$}
\drawloop[loopangle=180](4){$c$}
\drawloop[loopangle=135](5){$c$}
\drawloop[loopangle=0](1x){$a,c$}
\drawloop[loopangle=90](6){$c$}
\drawloop[loopangle=90](7){$c$}
\drawloop[loopangle=90](8){$c$}
\drawloop[loopangle=90](9){$c$}
\drawloop[loopangle=90](9x){$a,c$}
\end{picture}\end{center}
\caption{An irreducibly synchronizing strongly connected ternary automaton with $12$ states and reset length $95$.}\label{fig:gap12_95}
\end{figure}

There is no binary strongly connected automaton with $12$ states and reset length $95$, but we have constructed such an automaton over a ternary alphabet (Fig.~\ref{fig:gap12_95}). Similarly, we know an automaton for $n=9$ with reset length $53$ (second gap), and for $n=11$ with reset length $79$ (third gap).
This shows that the gaps, except the first one, are not necessarily preserved over larger alphabets.

\subsubsection{Extending words in one-cluster automata.}

One-cluster automata are an important class of synchronizing automata for which a quadratic bound on reset length has been found \cite{BBP2011QuadraticUpperBoundInOneCluster,Steinberg2011OneClusterPrime}.

Despite several attempts \cite{BBP2011QuadraticUpperBoundInOneCluster,CarpiDAlessandro2013IndependendSetsOfWords,Dubuc1998,Steinberg2011AveragingTrick,Steinberg2011OneClusterPrime}
at improving the bounds, so far, the \v{C}ern\'{y} conjecture has been proved only for one-cluster automata with a cycle of length $n$ (circular automata) or with a prime-length cycle. 
In \cite{Steinberg2011OneClusterPrime} an algebraic argument making use of ascending chain of linear subspaces and averaging trick has been applied. The proof is based on the claim that any subset $S \subset C$ on the cycle $C$ can be extended on this cycle by a word of length at most $\ell+n$ (we apply here the notation of Section~2). It is demonstrated that this holds in the case of prime length of $C$. Proving it for non-prime lengths would provide the proof of the \v{C}ern\'{y} conjecture for the whole class of one-cluster automata.

We have exhaustively searched for small examples of one-cluster automata with a non-prime cycle length such that the length $\ell+n$ of extending words is exceeded for some subset $S$, but found out that $\ell+n$ is sufficient in all tested cases, instead of the value $n+\ell+|C|-D^*(|C|,|S|)$ used to prove the bound from Theorem~\ref{thm:one-cluster_bound}.
Also, we found out that we can always use an extending word of the form $wa^{\ell}$ with $|w| \le n$, which is the form used in the proof for prime $|C|$.
\begin{conjecture}\label{con:one-cluster}
Let $\mathcal{A}$ be a one-cluster synchronizing automaton with a one-cluster letter $a$ with the cycle $C$ and level $\ell$.
For any non-empty proper subset $S \subset C$ there is a word $w$ such that $|S(wa^{\ell})^{-1} \cap C| > |S|$ and $|w| \le n$.
\end{conjecture}

In all the cases tested, for any $\ell$, non-prime $|C| < n$, and $|S|$ with $1 \le |S| < |C|$, we found an automaton for which we needed a word $w$ of length exactly $n$. 
So, it seems that the bound $|w| \le n$ is tight.

\subsubsection{Worst cases for the greedy compressing algorithm.}

The \emph{greedy compressing algorithm} is a well known approach for finding a reset word \cite{Ep1990,Pin1983OnTwoCombinatorialProblems,Volkov2008Survey}.
It starts from $S = Q$, and iteratively finds a shortest word $w$ such that $|Sw| < |S|$ and uses $Sw$ for next iteration, until $|S|=1$.
The concatenated words $w$ form the found reset word.
The length of the resulted reset word can vary, since there is ambiguity in selection of shortest words $w$.
By bounding the length of the found reset word we also obtain an upper bound for the reset length,
and in fact, the upper bound $(n^3-n)/6$ for the reset length is obtained by bounding the lengths of words $w$ for $|S|=2,\ldots,n$ and summing these bounds \cite{Pin1983OnTwoCombinatorialProblems}.
It is known that this algorithm finds a word of length $\Omega(n^2 \log n)$ for the \v{C}ern\'{y} automaton \cite{KariVolkov2013Handbook}, but it was not clear whether it is the worst case example.

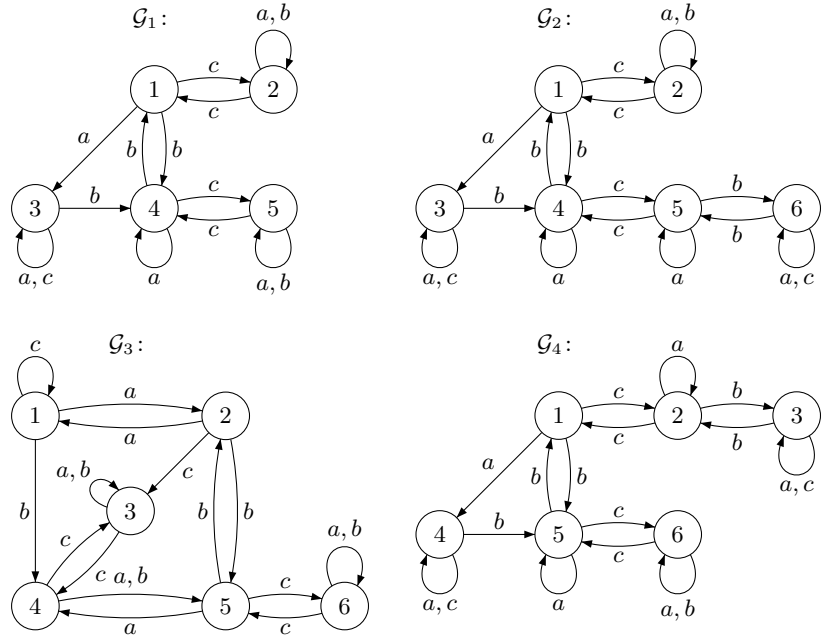
\begin{figure}[ht]
\unitlength 4.5pt\footnotesize
\begin{center}\begin{picture}(36,21)(-6,-5)
\gasset{Nh=4,Nw=4,Nmr=2,ELdist=.5,loopdiam=3}
\node[Nframe=n](name)(10,16){$\mathcal{G}_1\colon$}
\node(1)(10,10){$1$}
\node(2)(0,0){$3$}
\node(3)(10,0){$4$}
\node(4)(20,10){$2$}
\node(5)(20,0){$5$}
\drawedge[ELdist=-2](1,2){$a$}
\drawedge(2,3){$b$}
\drawedge[curvedepth=1](1,3){$b$}
\drawedge[curvedepth=1](3,1){$b$}
\drawloop[loopangle=270](2){$a,c$}
\drawloop[loopangle=270](3){$a$}
\drawedge[curvedepth=1](1,4){$c$}
\drawedge[curvedepth=1](4,1){$c$}
\drawedge[curvedepth=1](3,5){$c$}
\drawedge[curvedepth=1](5,3){$c$}
\drawloop(4){$a,b$}
\drawloop[loopangle=270](5){$a,b$}
\end{picture}\begin{picture}(32,21)(-4,-5)
\gasset{Nh=4,Nw=4,Nmr=2,ELdist=.5,loopdiam=3}
\node[Nframe=n](name)(10,16){$\mathcal{G}_2\colon$}
\node(1)(10,10){$1$}
\node(2)(0,0){$3$}
\node(3)(10,0){$4$}
\node(4)(20,10){$2$}
\node(5)(20,0){$5$}
\node(6)(30,0){$6$}
\drawedge[ELdist=-2](1,2){$a$}
\drawedge(2,3){$b$}
\drawedge[curvedepth=1](1,3){$b$}
\drawedge[curvedepth=1](3,1){$b$}
\drawloop[loopangle=270](2){$a,c$}
\drawloop[loopangle=270](3){$a$}
\drawedge[curvedepth=1](1,4){$c$}
\drawedge[curvedepth=1](4,1){$c$}
\drawedge[curvedepth=1](3,5){$c$}
\drawedge[curvedepth=1](5,3){$c$}
\drawloop(4){$a,b$}
\drawloop[loopangle=270](5){$a$}
\drawedge[curvedepth=1](5,6){$b$}
\drawedge[curvedepth=1](6,5){$b$}
\drawloop[loopangle=270](6){$a,c$}
\end{picture}\end{center}
\begin{center}\begin{picture}(36,28)(-6,-2)
\gasset{Nh=4,Nw=4,Nmr=2,ELdist=.5,loopdiam=3}
\node[Nframe=n](name)(8,22){$\mathcal{G}_3\colon$}
\node(1)(0,16){$1$}
\node(2)(16,16){$2$}
\node(3)(8,8){$3$}
\node(4)(0,0){$4$}
\node(5)(16,0){$5$}
\node(6)(26,0){$6$}
\drawedge[curvedepth=1](1,2){$a$}
\drawedge[curvedepth=1](2,1){$a$}
\drawedge[curvedepth=1](4,5){$a,b$}
\drawedge[curvedepth=1](5,4){$a$}
\drawloop[loopdiam=2,loopangle=145,ELdist=.2](3){$a,b$}
\drawloop(6){$a,b$}
\drawedge[ELside=r](1,4){$b$}
\drawedge[curvedepth=1](2,5){$b$}
\drawedge[curvedepth=1](5,2){$b$}
\drawedge(2,3){$c$}
\drawedge[curvedepth=1](3,4){$c$}
\drawedge[curvedepth=1](4,3){$c$}
\drawedge[curvedepth=1](5,6){$c$}
\drawedge[curvedepth=1](6,5){$c$}
\drawloop(1){$c$}
\end{picture}\begin{picture}(32,28)(-4,-8)
\gasset{Nh=4,Nw=4,Nmr=2,ELdist=.5,loopdiam=3}
\node[Nframe=n](name)(10,16){$\mathcal{G}_4\colon$}
\node(1)(10,10){$1$}
\node(2)(0,0){$4$}
\node(3)(10,0){$5$}
\node(4)(20,10){$2$}
\node(5)(20,0){$6$}
\node(6)(30,10){$3$}
\drawedge[ELdist=-2](1,2){$a$}
\drawedge(2,3){$b$}
\drawedge[curvedepth=1](1,3){$b$}
\drawedge[curvedepth=1](3,1){$b$}
\drawloop[loopangle=270](2){$a,c$}
\drawloop[loopangle=270](3){$a$}
\drawedge[curvedepth=1](1,4){$c$}
\drawedge[curvedepth=1](4,1){$c$}
\drawedge[curvedepth=1](3,5){$c$}
\drawedge[curvedepth=1](5,3){$c$}
\drawloop(4){$a$}
\drawloop[loopangle=270](5){$a,b$}
\drawedge[curvedepth=1](4,6){$b$}
\drawedge[curvedepth=1](6,4){$b$}
\drawloop[loopangle=270](6){$a,c$}
\end{picture}\end{center}
\caption{Automata $\mathcal{G}_1$, $\mathcal{G}_2$, $\mathcal{G}_3$, and $\mathcal{G}_4$, with the worst case length $19$, $30$, $28$, and $28$, and reset lengths $15$, $22$, $20$, and $20$, respectively. }\label{fig:worst_case_length}
\end{figure}

We experimentally tested the greedy algorithm for the worst cases.
Here, we restricted the studied class to irreducibly synchronizing automata, as otherwise we would get a lot of trivial examples derived from automata over a smaller alphabet.
By the \emph{worst case length} we mean the maximum length of the found word by the algorithm over all selections of shortest compressing words that can be taken by the algorithm.
For example, for automaton $\mathcal{G}_1$ from Fig.~\ref{fig:worst_case_length}, the worst case length is $19$ and a sequence of subsets considered by the greedy algorithm in the worst case can be the following:
$$Q = \{1,2,3,4,5\} \stackrel{b}{\longrightarrow} \{1,2,4,5\} \stackrel{aca}{\longrightarrow} \{3,4,5\} \stackrel{bcbacb}{\longrightarrow} \{1,4\} \stackrel{acbbcbaca}{\longrightarrow} \{3\}.$$
While this requires potentially very expensive computation, the worst case length can be computed by a kind of dynamic algorithm and $n-1$ iterations of breadth-first search in the power automaton.

It may be surprising that the \v{C}ern\'{y} automata generally do not exhibit the worst case length.
We have observed that for some values $n \ge 10$ the slowly synchronizing series $\mathrsfs{W}_n$, $\mathrsfs{D}''_n$, and $\mathrsfs{G}_n$ (see~\cite{AGV2010,AGV2013}) exceed the worst case length of the \v{C}ern\'{y} automaton with the same number of states.
In addition, we have found out four particular ternary examples shown in Fig.~\ref{fig:worst_case_length} exceeding the worst case length of the \v{C}ern\'{y} automaton with the same number of states, which do not seem generalizable to series.
Up to isomorphism, there are no more such examples within the range we have considered (Table~\ref{tab:conjectures}).

The results we have collected do not allow to state a reasonable conjecture.
So far, $\mathrsfs{W}_n$ is the best candidate for the largest worst case lengths for $n \ge 10$, and the \v{C}ern\'{y} automata for $n \le 9$, except $\mathcal{G}_1$ and $\mathcal{G}_2$ from Fig.~\ref{fig:worst_case_length} for $n=5,6$.
\begin{pbm}\label{pbm:worst_case_length}
What are the largest worst case lengths of the greedy compressing algorithm of automata with $n$ states?
\end{pbm}

It it noticeable that the dual \emph{greedy extending algorithm}, which starts from a singleton and uses shortest extending words rather than compressing ones, seem to have generally larger worst case lengths. For example, for the case of binary $n=7$ in the worst case it can find a reset word of length $48$ for some strongly connected automaton, whereas the greedy compressing algorithm finds a word of length at most $43$.

\subsubsection{Aperiodic synchronizing automata.}

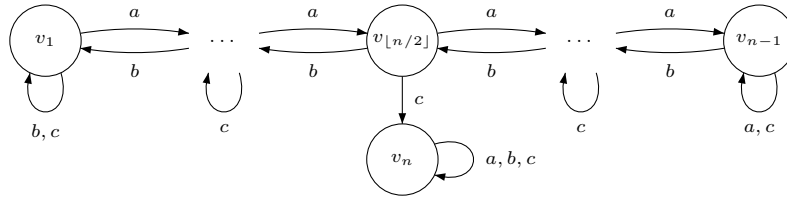
\begin{figure}[ht]
\unitlength 4.5pt\scriptsize
\begin{center}\begin{picture}(60,8.5)(0,0)
\gasset{Nh=6,Nw=6,Nmr=3,ELdist=1,loopdiam=3}
\node(1)(0,10){$v_1$}
\node[Nframe=n](dots1)(15,10){$\dots$}
\node(n/2)(30,10){$v_{\lfloor n/2 \rfloor}$}
\node[Nframe=n](dots2)(45,10){$\dots$}
\node(n-1)(60,10){$v_{n-1}$}
\node(n)(30,0){$v_n$}
\drawedge[curvedepth=1](1,dots1){$a$}
\drawedge[curvedepth=1](dots1,n/2){$a$}
\drawedge[curvedepth=1](n/2,dots2){$a$}
\drawedge[curvedepth=1](dots2,n-1){$a$}
\drawedge[curvedepth=1](n-1,dots2){$b$}
\drawedge[curvedepth=1](dots2,n/2){$b$}
\drawedge[curvedepth=1](n/2,dots1){$b$}
\drawedge[curvedepth=1](dots1,1){$b$}
\drawedge(n/2,n){$c$}
\drawloop[loopangle=270](1){$b,c$}
\drawloop[loopangle=270](n-1){$a,c$}
\drawloop[loopangle=0](n){$a,b,c$}
\drawloop[loopangle=270](dots1){$c$}
\drawloop[loopangle=270](dots2){$c$}
\end{picture}\end{center}
\caption{A ternary irreducibly synchronizing $n$-state aperiodic automaton with the reset length $n + \lfloor n/2 \rfloor - 2$.}\label{fig:aperiodic_ternary}
\end{figure}

Recall that an automaton is \emph{aperiodic} if there is no word inducing a transformation with a cycle of length $\ge 2$ (the transition semigroup has only trivial subgroups).
In~\cite{Volkov2008Survey} Volkov mentioned that although a quadratic upper bound for the reset length of aperiodic synchronizing automata has been proved, the largest reset length for known aperiodic automata does not exceed $n + \lfloor n/2 \rfloor -2$.
This length is reached by a series of binary automata constructed by Ananichev \cite{An2005MortalityThresholdForPartiallyMonotonic}.
In this connection, it may be interesting to know that the same bound is also reached for every $n > 1$ by a series of irreducibly ternary aperiodic automata. It has a quite simple definition and an easy proof for the reset length (comparing with~\cite{An2005MortalityThresholdForPartiallyMonotonic}).
Let $\mathcal{A}_n = \langle Q,\{a,b,c\},\delta\rangle$, where $Q = \{v_1,\ldots,v_n\}$, $\delta(v_i,a)=v_{i+1}$ for $1\le i \le n-2$, $\delta(v_i,b)=v_{i-1}$ for $2 \le i \le n-1$, $\delta(v_{\lfloor n/2\rfloor},c)=v_n$, and $\delta(v_i,x)=v_i$, otherwise ($x\in\Sigma$) (shown in~Fig.~\ref{fig:aperiodic_ternary}).

Volkov\footnote{personal communication} has also pointed out that $n-1$ may be an upper bound for the reset length in the class of \emph{strongly connected} synchronizing aperiodic automata, but there was not enough evidence.
The bound can be met trivially if the underlying digraph of the automaton is a \emph{bidirectional path}:
$Q = 1,\ldots,n$, for every $1 \le i \le n-1$ there are the directed edges $(i,i+1)$ and $(i+1,i)$, and every edge that is not a loop is of that form.

Since our verifications involve a huge number of aperiodic automata, we experimentally support the following conjectures:
\begin{conjecture}[cf.\ \textnormal{\cite{Volkov2008Survey}}]\label{con:aperiodic}
Every synchronizing aperiodic automaton with $n>1$ states has a reset word of length at most $\leq n + \lceil n/2 \rceil - 2$.
\end{conjecture}
\begin{conjecture}[Volkov]\label{con:strongly_aperiodic}
Every strongly connected synchronizing automaton has a reset word of length at most $n-1$.
Moreover, if this bound is met, then the underlying digraph of the automaton is a bidirectional path.
\end{conjecture}

\subsubsection{Avoiding states.}

In a recent short note \cite{GJT2014ANoteOnARecentAttempt} the authors state the following problem related to the recent unsuccessful attempt of improving the general upper bound on reset length \cite{Tr2011ModifyingUpperBound}:  
Given a strongly connected synchronizing automaton, what is the minimal length $\ell$ such that for any $q \in Q$ there is a word $w$ of length $\le \ell$ and such that $q \not\in Qw$. If $\ell \in O(n)$, then we would obtain a better upper bound than $(n^3-n)/6$.

Experimentally, we have found out what the value of $\ell$ for a given $n$ might be, and provided support for the following conjecture:
\begin{conjecture}\label{con:avoiding_states}
In a synchronizing strongly connected automaton, for any $q \in Q$ there is a word $w$ of length $\le 2n-2$ and such that $q \not\in Qw$.
This bound is tight for $n \ge 4$ over a ternary alphabet.
\end{conjecture}
Recently, Vojt\v{e}ch Vorel\footnote{personal communication, unpublished} discovered an infinite series of binary automata whose minimal length in question is $2n-4$, which is currently the best theoretical lower bound for the problem.

\subsubsection{New rank conjecture.}

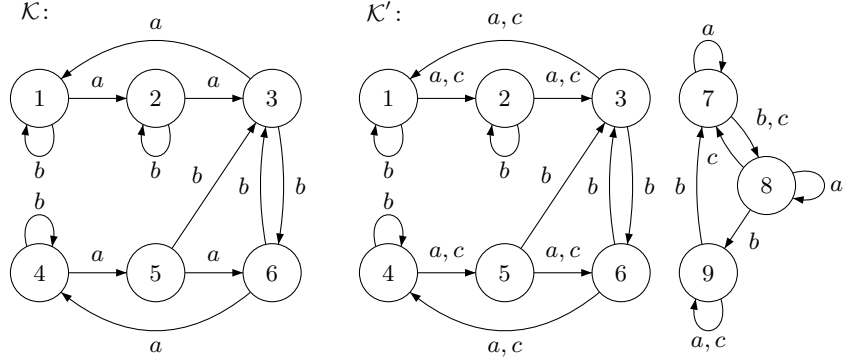
\begin{figure}
\unitlength 22pt
\begin{center}\begin{picture}(6,5.5)(0,0)
\gasset{Nh=1,Nw=1,Nmr=1,ELdist=0.2,loopdiam=0.5}
\node[Nframe=n](name)(0,5.5){$\mathcal{K}\colon$}
\node(v1)(0,4){$1$}
\node(v2)(2,4){$2$}
\node(v3)(4,4){$3$}
\node(v4)(0,1){$4$}
\node(v5)(2,1){$5$}
\node(v6)(4,1){$6$}
\drawedge(v1,v2){$a$}
\drawedge(v2,v3){$a$}
\drawedge[curvedepth=-1,ELside=r](v3,v1){$a$}
\drawedge(v4,v5){$a$}
\drawedge(v5,v6){$a$}
\drawedge[curvedepth=1](v6,v4){$a$}
\drawloop[loopangle=-90,ELdist=0.1](v1){$b$}
\drawloop[loopangle=-90,ELdist=0.1](v2){$b$}
\drawedge[curvedepth=0.2](v3,v6){$b$}
\drawloop[loopangle=90,ELdist=0.1](v4){$b$}
\drawedge(v5,v3){$b$}
\drawedge[curvedepth=0.2](v6,v3){$b$}
\end{picture}\begin{picture}(7,5.5)(0,0)
\gasset{Nh=1,Nw=1,Nmr=1,ELdist=0.2,loopdiam=0.5}
\node[Nframe=n](name)(0,5.5){$\mathcal{K}'\colon$}
\node(v1)(0,4){$1$}
\node(v2)(2,4){$2$}
\node(v3)(4,4){$3$}
\node(v4)(0,1){$4$}
\node(v5)(2,1){$5$}
\node(v6)(4,1){$6$}
\drawedge(v1,v2){$a,c$}
\drawedge(v2,v3){$a,c$}
\drawedge[curvedepth=-1,ELside=r](v3,v1){$a,c$}
\drawedge(v4,v5){$a,c$}
\drawedge(v5,v6){$a,c$}
\drawedge[curvedepth=1](v6,v4){$a,c$}
\drawloop[loopangle=-90,ELdist=0.1](v1){$b$}
\drawloop[loopangle=-90,ELdist=0.1](v2){$b$}
\drawedge[curvedepth=0.2](v3,v6){$b$}
\drawloop[loopangle=90,ELdist=0.1](v4){$b$}
\drawedge(v5,v3){$b$}
\drawedge[curvedepth=0.2](v6,v3){$b$}
\node(v7)(5.5,4){$7$}
\node(v8)(6.5,2.5){$8$}
\node(v9)(5.5,1){$9$}
\drawedge[curvedepth=0.2](v7,v8){$b,c$}
\drawedge(v8,v9){$b$}
\drawedge[curvedepth=0.2](v9,v7){$b$}
\drawedge[curvedepth=0.2](v8,v7){$c$}
\drawloop[loopangle=90,ELdist=0.1](v7){$a$}
\drawloop[loopangle=0,ELdist=0.1](v8){$a$}
\drawloop[loopangle=-90,ELdist=0.1](v9){$a,c$}
\end{picture}\end{center}
\caption{The Kari automaton $\mathcal{K}$ \cite{Kari2001Counterexample}, and a Kari-like automaton $\mathcal{K}'$.}\label{fig:kari_automaton}
\end{figure}

Pin~\cite{Pin1983OnTwoCombinatorialProblems} proposed the following generalization of the \v{C}ern\'{y} conjecture: For every 
$0<d, n$, if there is a word of rank $\le n-d$, then there is such a word of length $\le d^2$. Pin proved this for $d \le 3$. However, Kari~\cite{Kari2001Counterexample} found a celebrated counterexample to this conjecture for $d=4$, which is a binary automaton $\mathcal{K}$ with $6$ states (Fig.~\ref{fig:kari_automaton}).
As a consequence, a modification of this generalized conjecture was proposed restricting it to $d$ being the rank of the considered automaton (see for example \cite{AS2009MatrixMortalityAndCernyPin}). However this seems to be a quite radical restriction. 

In our computations, we have found no other counterexample to Pin's conjecture except for trivial extensions and modifications. This may suggest that Kari construction works due to the number of involved states small enough, and is, in fact, an exception. By a \emph{trivial extension} of an automaton over alphabet $\Sigma$ we mean one obtained by adding letters to $\Sigma$ that acts either as the identity transformation or as any letter in $\Sigma$. So a trivial extension has the same number of the states and the transition semigroup, and trivial extensions of the Kari automaton $\mathcal{K}$ are counterexamples to the Pin's conjecture, for $d=4$, as well. By a \emph{disjoint union} of two automata $\mathcal{A} = \langle Q, \Sigma, \delta\rangle$ and $\mathcal{A}' = \langle Q', \Sigma, \delta'\rangle$ we mean the construction where the automata have the same alphabet $\Sigma$, and disjoint sets of states $Q,Q'$, and the union is simply $\mathcal{A} = \langle Q\cup Q', \Sigma, \delta\cup\delta'\rangle$. If we take a disjoint union of $\mathcal{K}$ with any \emph{permutation automaton} (one whose letters act like permutations, or in other words, one of rank equal to its size), then again we get a counterexample to the Pin's conjecture, for $d=4$. Yet, in all these automata the failure is caused by the same Kari construction on the set of the 6 states. In our experiments, we have discovered no other counterexample. This may be treated as an evidence for the conjecture we state below.

Consider the smallest class of automata containing $\mathcal{K}$ and closed on taking trivial extension and disjoint union with permutation automata. Let us call automata in this class \emph{Kari-like} automata (see Fig.~\ref{fig:kari_automaton}). Then we have
\begin{conjecture}\label{con:kari-like}
For every $d$, if an automaton $\mathcal{A}$ has a word of rank at most $n-d$, then there is such a word of length at most $d^2$, unless $\mathcal{A}$ is a Kari-like automaton and $d=4$ (in which case there is a word of rank $n-4$ of length $d^2+1 = 17$). 
\end{conjecture}

\subsubsection{Subset synchronization.}

The last conjecture was posed by {\^A}ngela Cardoso:
\begin{conjecture}[Cardoso \textrm{\cite{Cardoso2014PhD}}]\label{con:subset_synchr}
In a synchronizing automaton, for any subset $S$ of states there is a word $w$ with $|Sw|=1$ of length at most
$$(n-1)^2-\left\lceil\frac{n-|S|}{|S|}\right\rceil\left(2n-|S|\left\lceil\frac{n}{|S|}\right\rceil-1\right).$$
\end{conjecture}
This is another generalization of the \v{C}ern\'{y} conjecture, and it can be viewed as a counterpart for the rank conjecture, where we bound the length of words compressing $Q$ to a subset of the given size, rather than a subset to a singleton.

Conjecture~\ref{con:subset_synchr} has been proved for several special classes of automata, and the formula is tight for any subset size in the \v{C}ern\'{y} series.
Besides confirmation for small automata, we identified 18 particular examples of irreducibly synchronizing automata with $n \in \{3,4,5,6\}$ states meeting the bound for some subset $S$ that are not isomorphic to the \v{C}ern\'{y} automata.
Note that the conjecture is not true in general for non-synchronizing automata, as Vorel \cite{Vorel2014SubsetSynchronizationOfTransitiveAutomata} has constructed a series of non-synchronizing strongly connected binary automata with subsets whose shortest synchronizing words are of exponential length.

\subsection{Summary}

Table~\ref{tab:conjectures} summarizes the ranges for which the discussed conjectures have been confirmed or the problems checked.
The ranges vary due to different numbers of automata that have to be checked, computational complexity of verification for a single automaton, and computation time devoted for each of the problems.

\renewcommand{\arraystretch}{1.2}
\begin{table}[ht]\centering
\newcolumntype{C}{>{\centering\let\newline\\\arraybackslash\hspace{0pt}}m{.52cm}}
\caption{Experimental verification of conjectures.
The numbers denote the size of the alphabet up to which the given conjecture has been checked.
The symbol $\infty$ denotes that the problem has been verified for all automata with the given number of states and any number of letters.}\label{tab:conjectures}
\begin{tabular}{|l|C|C|C|C|C|C|C|C|C|}\hline
\multirow{2}{*}{Problem}&\multicolumn{9}{|c|}{Number of states $n$} \\ \cline{2-10}
                                                                      &$\le 4$ & 5      & 6 & 7 & 8 & 9 & 10 & 11 & 12 \\ \hline\hline
\v{C}ern\'{y} conjecture and \cite[Conjecture~1]{AGV2013}             &$\infty$&$\infty$& 6 & 4 & 3 & 2 & 2  & 2  & 2  \\ \hline
Conjecture~\ref{con:one-cluster} (one-cluster)                        &$\infty$& 5      & 4 & 3 & 2 & 2 & 2  &    &    \\ \hline
Problem~\ref{pbm:worst_case_length} (greedy algorithm)                &$\infty$& 6      & 4 & 3 & 2 & 2 & 2  &    &    \\ \hline
Conjecture~\ref{con:aperiodic} (aperiodic)                            &$\infty$& 5      & 3 & 3 & 2 & 2 & 2  &    &    \\ \hline
Conjecture~\ref{con:strongly_aperiodic} (strongly connected aperiodic)&$\infty$& 8      & 5 & 3 & 2 & 2 & 2  & 2  &    \\ \hline
Conjecture~\ref{con:avoiding_states} (avoiding states)                &$\infty$& 8      & 4 & 3 & 2 & 2 & 2  &    &    \\ \hline
Conjecture~\ref{con:kari-like} (new rank conjecture)                  &$\infty$&$\infty$& 5 & 3 & 3 & 2 & 2  &    &    \\ \hline
Conjecture~\ref{con:subset_synchr} (subset synchronization)           &$\infty$&$\infty$& 5 & 4 & 3 & 2 & 2  &    &    \\ \hline
\end{tabular}
\end{table}
\vspace{-.9cm}


\subsubsection{Acknowledgments.}
We thank Mikhail Volkov for suggesting Conjecture~\ref{con:strongly_aperiodic},
and Mikhail Berlinkov for observing that the bound for one-cluster automata can be improved for periodic subsets on the cycle, which leaded to an improvement of our algorithm.
We thank also Vojt\v{e}ch Vorel for discussing the problem of avoiding states and sharing the series.
The main part of the computations was performed on a grid that belongs to Institute of Computer Science of Jagiellonian University (thanks to Adam Roman).

\bibliographystyle{splncs03}

\end{document}